\documentclass[11pt,a4paper]{article}

\usepackage{jcappub,bm}

\title{Galvano-rotational effect induced by electroweak
interactions in pulsars}

\author[a,b,c]{Maxim Dvornikov}

\affiliation[a]{Institute of Physics, University of S\~{a}o Paulo,
CP 66318, CEP 05314-970 S\~{a}o Paulo, SP, Brazil}
\affiliation[b]{Pushkov Institute of Terrestrial Magnetism, Ionosphere
and Radiowave Propagation (IZMIRAN),
142190 Troitsk, Moscow, Russia}
\affiliation[c]{Physics Faculty, National Research Tomsk State University,
36 Lenin Ave., 634050 Tomsk, Russia}
\emailAdd{maxdvo@izmiran.ru}

\abstract{
We study electroweakly interacting particles in rotating matter. The existence of the electric current along the axis of the matter rotation is predicted in this system.
This new galvano-rotational effect is caused by the parity violating interaction
between massless charged particles in the rotating matter. We start
with the exact solution of the Dirac equation for a fermion involved
in the electroweak interaction in the rotating frame. This equation
includes the noninertial effects. Then, using the obtained solution,
we derive the induced electric current which turns out to flow along
the rotation axis. We study the possibility of the appearance of the galvano-rotational effect in dense
matter of compact astrophysical objects. The particular
example of neutron and hypothetical quark stars is discussed. It is shown that,
using this effect, one can expect the generation of toroidal magnetic fields comparable
with poloidal ones in old millisecond pulsars. We also briefly discuss
the generation of the magnetic helicity in these stars. Finally we analyze the possibility to apply the galvano-rotational effect for the description of the asymmetric neutrino emission from a neutron star to explain pulsars kicks.
}

\keywords{neutron stars, magnetic fields, gravity}

\arxivnumber{1503.00608}

\begin{document}

\maketitle

\section{Introduction\label{sec:INTR}}

The importance of noninertial effects for various areas in modern
physics cannot be underestimated. Some of the examples of these effects
are mentioned in ref.~\cite{Ryd09}. One of the most common manifestations
of noninertial effects is the description of physical processes in
a rotating frame. One can expect the appearance of additional interesting phenomena if, besides the matter rotation, there is a parity violating interaction in the system. In the
present work we will show that, in this situation, an electric current flowing along the rotation axis can be induced. We shall
call this phenomenon as the new \emph{galvano-rotational effect}
(GRE).

Previously the generation of an electric current due to nontrivial
topological effects was studied mainly in connection to the chiral
magnetic effect (CME)~\cite{Kha15}. In that case a nonzero current can
be induced in the system of massless particles embedded in an external
magnetic field~\cite{Vil80}, provided there is an imbalance between
left and right particles. Recently, in ref.~\cite{DvoSem15}, we
showed that the electric current can be generated even at zero chiral
imbalance if charged particles are involved in the parity violating
interaction.
%Note that the analogy between the propagation in a rotating
%parity violating electroweak matter and the motion in an external
%magnetic field was previously mentioned in ref.~\cite{GriSavStu07}.

We will apply the new GRE in astrophysical media to generate a toroidal magnetic field (TMF) inside a compact star. Although
stellar TMFs cannot be observed directly, they are an internal ingredient
of various astrophysical objects. For example, the most plausible explanation
of the $22\thinspace\text{yr}$ solar cycle is the oscillation between poloidal and
toroidal components of the solar magnetic field~\cite{BraSub05}.
Moreover, purely poloidal or toroidal stellar magnetic fields were
shown in refs.~\cite{FloRud77,Tay73} to be unstable. Thus TMF is
inherent to a magnetized star. In this work we show how TMF can be
generated in an old millisecond pulsar basing on GRE.

Since macroscopic fluxes of electroweakly interacting particles are produced by GRE, we can try to apply this effect to explain linear velocities of millisecond pulsars. It is known from astronomical observations~\cite{Hob05} that pulsars possess great linear velocities. Nevertheless physical processes underlaying pulsar kicks are still unclear. It might be reasonable to use GRE to account for pulsars kicks due to, e.g., anisotropic neutrino emission since linear velocities of pulsars were reported in ref.~\cite{Joh05} to be correlated with their angular velocities.

This paper is organized in the following way. First, in section~\ref{sec:FLATST},
we briefly describe the Standard Model interaction between leptons
and quarks in flat space-time. Then, in section~\ref{sec:DIRACEQ},
we derive the Dirac equation for a fermion which interacts electroweakly
with a rotating background matter. For this purpose we write down
this Dirac equation in the corotating frame, using the method of an effective curved space-time. The exact solution of
the Dirac equation for an ultrarelativistic fermion, accounting for
the noninertial effects, is obtained in section~\ref{sec:DIRACEQ}.
In section~\ref{sec:CURRCALC}, we establish GRE, which, in
this situation, consists in the appearance of the electric current
along the rotation axis. We calculate this current in section~\ref{sec:CURRCALC}
using the exact solution of the Dirac equation obtained in section~\ref{sec:DIRACEQ}.
In section~\ref{sec:TMFGENERATION}, we apply GRE for the
generation of TMF and the magnetic helicity in compact rapidly rotating stars. Finally, in section~\ref{sec:KICKS}, we try to use GRE to produce anisotropic neutrino emission from pulsars to explain their great linear velocities.

In section~\ref{sec:CONCL},
we summarize our results and compare them with the findings of other
authors. Some details of the derivation of the electric current in rotating matter are provided in appendix~\ref{sec:CURRDER}.

\section{Electroweak interaction of fermions in flat space-time\label{sec:FLATST}}

In this section we shall briefly remind the description of the electroweak
interaction between leptons and quarks in flat space-time.

Let us consider a medium consisting of electrons, electron neutrinos $\nu_e$ as well as $u$ and
$d$ quarks. We shall assume that quarks are both in confined states, forming
nucleons, and hypothetical free particles. The effective Lagrangian
for the electroweak interaction in this system in the Fermi approximation
has form~\cite{MohPal04},
\begin{equation}\label{eq:LFermi}
  \mathcal{L}_{\mathrm{eff}} = -\frac{4G_{\mathrm{F}}}{\sqrt{2}}
  \left[
    J^{\mu}J_{\mu}^{\dagger}+K^{\mu}K_{\mu}
  \right],
\end{equation}
where $G_{\mathrm{F}}\approx1.17\times10^{-5}\thinspace\text{GeV}^{-2}$
is the Fermi constant, $J^{\mu}$ is the charged current, and $K^{\mu}$
is the neutral current.

In the considered system of elementary particles, the charged current
has the form,
\begin{equation}\label{eq:cc}
  J_{\mu} = V_{ud}\bar{\psi}_{u}\gamma_{\mu}^{\mathrm{L}}\psi_{d} +
  \bar{\psi}_{\nu_e}\gamma_{\mu}^{\mathrm{L}}\psi_{e},
\end{equation}
where $\psi_{u,d}$ are the wave functions of $u$ and $d$ quarks, $\psi_{\nu_e, e}$ are the wave functions of ${\nu_e}$ and an electron,
$\gamma_{\mu}^{\mathrm{L}}=\gamma_{\mu}\left(1-\gamma^{5}\right)/2$,
$\gamma^{\mu}=\left(\gamma^{0},\bm{\gamma}\right)$ are the Dirac
matrices, $\gamma^{5}=\mathrm{i}\gamma^{0}\gamma^{1}\gamma^{2}\gamma^{3}$,
and $V_{ud}\approx0.97$ is the element of the Cabbibo-Kobayashi-Maskawa
matrix. The neutral current can be expressed as
\begin{equation}\label{eq:nc}
  K_{\mu} = \sum_{f=u,d,e,\nu_e}
  \left[
    \epsilon_{f}^{\mathrm{L}}\bar{\psi}_{f}\gamma_{\mu}^{\mathrm{L}}\psi_{f} +
    \epsilon_{f}^{\mathrm{R}}\bar{\psi}_{f}\gamma_{\mu}^{\mathrm{R}}\psi_{f}
  \right],
\end{equation}
where $\gamma_{\mu}^{\mathrm{R}}=\gamma_{\mu}\left(1+\gamma^{5}\right)/2$
and
\begin{alignat}{4}
  & \epsilon_{e}^{\mathrm{L}} = -\frac{1}{2}+\xi,
  \quad &&
  \epsilon_{\nu_e}^{\mathrm{L}} = \frac{1}{2},
  \quad &&
  \epsilon_{u}^{\mathrm{L}} = \frac{1}{2}-\frac{2}{3}\xi,
  \quad &&
  \epsilon_{d}^{\mathrm{L}}= - \frac{1}{2}+\frac{1}{3}\xi,
  \nonumber
  \\
  & \epsilon_{e}^{\mathrm{R}} = \xi,
  \quad &&
  \epsilon_{\nu_e}^{\mathrm{R}} = 0,
  \quad &&
  \epsilon_{u}^{\mathrm{R}} = -\frac{2}{3}\xi,
  \quad &&
  \epsilon_{d}^{\mathrm{R}} = \frac{1}{3}\xi.
\end{alignat}
%
%\begin{align}
%  \epsilon_{e}^{\mathrm{L}}= & -\frac{1}{2}+\xi,
%  \quad\epsilon_{e}^{\mathrm{R}}=\xi,\nonumber \\
%\epsilon_{u}^{\mathrm{L}}= & \frac{1}{2}-\frac{2}{3}\xi,\quad\epsilon_{u}^{\mathrm{R}}=-\frac{2}{3}\xi,\nonumber \\
%\epsilon_{d}^{\mathrm{L}}= & \frac{1}{2}+\frac{1}{3}\xi,\quad\epsilon_{d}^{\mathrm{R}}=\frac{1}{3}\xi.
%\end{align}
Here $\xi=\sin^{2}\theta_{\mathrm{W}}\approx0.23$ and $\theta_{\mathrm{W}}$
is the Weinberg angle.

In this section we shall consider the case when background fermions
are at rest and unpolarized. While averaging the currents in eqs.~(\ref{eq:cc})
and~(\ref{eq:nc}) over the Fermi-Dirac distributions $\left\langle \dots\right\rangle $,
in this case we get that only $\left\langle \psi_{f}^{\dagger}\psi_{f}\right\rangle =n_{f}\neq0$,
where $n_{f}$ is the invariant number density of these fermions. The quantities
$\left\langle \bar{\psi}_{f}\bm{\gamma}\psi_{f} \right\rangle = 0$
and $\left\langle \bar{\psi}_{f}\gamma^{\mu}\gamma^{5}\psi_{f}\right\rangle = 0$
are vanishing since they are proportional to the macroscopic velocity
and the polarization.

After averaging over the ensemble of background particles, we can rewrite
eq.~(\ref{eq:LFermi}) in the form,
\begin{equation}\label{eq:Lenp}
  \mathcal{L}_{\mathrm{eff}} = -\bar{\psi}
  \left[
    \gamma_{0}^{\mathrm{L}}V_{\mathrm{L}}+\gamma_{0}^{\mathrm{R}}V_{\mathrm{R}}
  \right]
  \psi,
\end{equation}
where $\psi$ is the wave function of a test fermion which undergoes
a scattering off background particles and the effective potentials
$V_{\mathrm{L,R}}$ are given in table~\ref{tab:VLR} for any scattering
channels.

When we consider the electron and $\nu_e$ scattering off nucleons, $u$
and $d$ quarks are confined inside neutrons or protons. In the case of
electron-nucleons interaction, only the neutral current contributes to eq.~(\ref{eq:Lenp}). If
we study the scattering $u$ quarks off $d$ quarks, and vice versa, as well as the $\nu_e$ scattering off background fermions,
both the charged and the neutral currents give the contributions to
eq.~(\ref{eq:Lenp}). To derive the expression for $V_{\mathrm{L}}$
for $(ud)$, $(du)$, and $(\nu_e e)$ interactions on the basis of eq.~(\ref{eq:cc}),
we use the Fierz transformation. We also note that we consider the $\nu_e$ interaction with electroneutral matter where $n_e = n_p$.

\begin{table}
  \centering
  \begin{tabular}{|p{1.5cm}|p{2cm}|c|c|}
    \hline
    Test \ \ \ \ \ \ \ \ particle & Background particles & $V_{\mathrm{L}}$ & $V_{\mathrm{R}}$
    \tabularnewline
    \hline
    \hline
    electron & nucleons &
    $-\frac{G_{\mathrm{F}}}{\sqrt{2}}
    \left[
      n_{n}-n_{p}(1-4\xi)
    \right]
    (2\xi-1)$ & $-\frac{G_{\mathrm{F}}}{\sqrt{2}}
    \left[
      n_{n}-n_{p}(1-4\xi)
    \right]2\xi$
    \tabularnewline
    \hline
    $\nu_e$ & nucleons \& electrons &
    $-\frac{G_{\mathrm{F}}}{\sqrt{2}}
    \left[
      n_{n}- 2 n_{e}
    \right]$ &
    0
    \tabularnewline
    \hline
    $u$ quark & $d$ quarks & $ - \frac{G_{\mathrm{F}}}{\sqrt{2}}n_{d}
    \left(
      1 - \frac{8}{3}\xi + \frac{16}{9}\xi^{2}-2|V_{ud}|^{2}
    \right)$
    &
    $\frac{G_{\mathrm{F}}}{\sqrt{2}}n_{d}
    \left(
      \frac{4}{3}\xi-\frac{16}{9}\xi^{2}
    \right)$
    \tabularnewline
    \hline
    $d$ quark & $u$ quarks & $-\frac{G_{\mathrm{F}}}{\sqrt{2}}n_{u}
    \left(
      1 - \frac{10}{3}\xi+\frac{16}{9}\xi^{2}-2|V_{ud}|^{2}
    \right)$
    &
    $\frac{G_{\mathrm{F}}}{\sqrt{2}}n_{u}
    \left(
      \frac{2}{3}\xi-\frac{16}{9}\xi^{2}
    \right)$
    \tabularnewline
    \hline
  \end{tabular}
  \protect\caption{The values of the effective potentials $V_{\mathrm{L,R}}$ in eq.~(\ref{eq:Lenp})
  for various channels of the scattering of a test fermion off background
  particles. Here $n_e$ is the electron density, $n_{n,p}$ are the densities of neutrons and protons,
  and $n_{u,d}$ are the densities of $u$ and $d$ quarks.\label{tab:VLR}}
\end{table}

\section{Dirac equation for a fermion interacting with rotating matter\label{sec:DIRACEQ}}

In this section we shall find the exact solution of the Dirac equation
for a fermion interacting with a rotating matter by means of the electroweak
forces. The obtained solution includes noninertial effects.

Let us discuss the interaction of a fermion with matter rotating with
the constant angular velocity $\omega$. In this case we cannot directly
apply the results of section~\ref{sec:FLATST} taking $\left\langle \bar{\psi}_{f}\bm{\gamma}\psi_{f}\right\rangle \sim\mathbf{v}_{f}\neq0$,
where $\mathbf{v}_{f}=(\bm{\omega}\times\mathbf{r})$ is the fermions
velocity, while averaging over the ensemble of background particles.
In the situation, when matter moves with an acceleration, one should
account for possible noninertial effects.

Nevertheless we can still choose a noninertial reference frame where
matter is at rest. Assuming that matter is unpolarized, we get that
only $\left\langle \psi_{f}^{\dagger}\psi_{f}\right\rangle \neq0$
in this reference frame. Thus, formally we can use the effective potentials
derived in section~\ref{sec:FLATST}. For the first time this approach
was put forward in ref.~\cite{Dvo14}, where the neutrino interaction
with a rotating matter was studied.

It is known that the description of a particle in an accelerated frame
is analogous to the motion of this particle in the curved space-time
or the interaction with an effective gravitational field. For example,
when we study the motion in the rotating frame, the interval takes
the form~\cite{LanLif94},
\begin{equation}\label{eq:mertrot}
  \mathrm{d}s^{2} =
  g_{\mu\nu}\mathrm{d}x^{\mu}\mathrm{d}x^{\nu} =
  (1-\omega^{2}r^{2})\mathrm{d}t^{2}-\mathrm{d}r^{2}-2\omega r^{2}\mathrm{d}t\mathrm{d}\phi-r^{2}\mathrm{d}\phi^{2}-\mathrm{d}z^{2},
\end{equation}
where $g_{\mu\nu}$ is the metric tensor of the effective gravitational
field. Here we use the cylindrical coordinates $x^{\mu}=(t,r,\phi,z)$.

Using eq.~(\ref{eq:Lenp}), we get that the Dirac equation for a
test fermion, with the mass $m$, involved in the parity violating
interaction and moving in a curved space-time, has the form (see also
ref.~\cite{Dvo14}),
\begin{equation}\label{eq:Depsicurv}
  \left[
    \mathrm{i}\gamma^{\mu}(x)\nabla_{\mu}-m
  \right]
  \psi =
  \gamma_{\mu}(x)
  \left\{
    \frac{V_{\mathrm{L}}^{\mu}}{2}
    \left[
      1-\gamma^{5}(x)
    \right] +
    \frac{V_{\mathrm{R}}^{\mu}}{2}
    \left[
      1+\gamma^{5}(x)
    \right]
  \right\}
  \psi,
\end{equation}
where $\gamma^{\mu}(x)$ are the coordinate dependent Dirac matrices,
$\nabla_{\mu}=\partial_{\mu}+\Gamma_{\mu}$ is the covariant derivative,
$\Gamma_{\mu}$ is the spin connection, $\gamma^{5}(x) = - (\mathrm{i}/4!) E^{\mu\nu\alpha\beta} \gamma_{\mu}(x)\gamma_{\nu}(x)\gamma_{\alpha}(x)\gamma_{\beta}(x)$,
$E^{\mu\nu\alpha\beta} = \varepsilon^{\mu\nu\alpha\beta} / \sqrt{-g}$
is the covariant antisymmetric tensor in curved space-time, $g=\det(g_{\mu\nu})$,
and $V_{\mathrm{L,R}}^{\mu}$ are the effective potentials. Note that,
since we choose a corotating frame, then $V_{\mathrm{L,R}}^{0}\equiv V_{\mathrm{L,R}}\neq0$
and $V_{\mathrm{L,R}}^{i}=0$, where $V_{\mathrm{L,R}}$ are given
in table~\ref{tab:VLR}. Analogous Dirac equation was also discussed in
ref.~\cite{PirRoyWud96}.

One can check that, using the following vierbein vectors:
\begin{equation}\label{eq:vierbein}
  e_{0}^{\ \mu} = (1,0,-\omega,0),
  \quad
  e_{1}^{\ \mu} = (0,1,0,0),
  \quad
  e_{2}^{\ \mu} = (0,0,1/r,0),
  \quad
  e_{3}^{\ \mu} = (0,0,0,1),
\end{equation}
the metric tensor in eq.~(\ref{eq:mertrot}) can be diagonalized,
$\eta_{ab} = e_{a}^{\ \mu}e_{b}^{\ \nu}g_{\mu\nu}$, where $\eta_{ab} = \text{diag}(1, -1, \\ -1, -1)$
is the metric in a locally Minkowskian frame.

Let us introduce the constant Dirac matrices in a locally Minkowskian
frame by $\gamma^{\bar{a}}=e_{\ \mu}^{a}\gamma^{\mu}(x)$. As shown
in ref.~\cite{Dvo14}, $\gamma^{5}(x) = \mathrm{i}\gamma^{\bar{0}}\gamma^{\bar{1}}\gamma^{\bar{2}}\gamma^{\bar{3}} = \gamma^{\bar{5}}$
does not depend on coordinates. The spin connection in the Dirac eq.~(\ref{eq:Depsicurv})
has the form~\cite{GriMamMos80},
\begin{equation}\label{eq:spincon}
  \Gamma_{\mu} =
  -\frac{\mathrm{i}}{4}\sigma^{ab}\omega_{ab\mu},
  \quad
  \omega_{ab\mu} = e_{a}^{\ \nu}e_{b\nu;\mu},
\end{equation}
where $\sigma_{ab} = (\mathrm{i}/{2})[\gamma_{\bar{a}},\gamma_{\bar{b}}]_{-}$ are
the generators of the Lorentz transformations in a locally Minkowskian
frame and the semicolon stays for the covariant derivative. The explicit
calculation on the basis of eq.~(\ref{eq:spincon}) shows that the
nonzero components of the connection one-form $\omega_{ab} = \omega_{ab\mu}\mathrm{d}x^{\mu}$
are
\begin{equation}\label{eq:spinconcomp}
  \omega_{12\mu} = -\omega_{21\mu}=(\omega,0,1,0).
\end{equation}
Using eqs.~(\ref{eq:spincon}) and~(\ref{eq:spinconcomp}) we get
that $\mathrm{i}\gamma^{\mu}(x)\Gamma_{\mu}=\mathrm{i} \gamma^{\bar{1}}/2r$.

Using the definition of $\gamma^{\bar{a}}$, the Dirac eq.~(\ref{eq:Depsicurv})
takes the form,
\begin{align}\label{eq:Direqgen}
  \mathcal{D}\psi= &
  \left(
    \gamma^{\bar{0}}-\omega r\gamma^{\bar{2}}
  \right)
  \left(
    V_\mathrm{V}-V_\mathrm{A}\gamma^{\bar{5}}
  \right)\psi,
  \nonumber
  \\
  \mathcal{D}= &
  \left[
    \mathrm{i}\gamma^{\bar{0}}
    \left(
      \partial_{0}-\omega\partial_{\phi}
    \right) +
    \mathrm{i}\gamma^{\bar{1}}
    \left(
      \partial_{r}+\frac{1}{2r}
    \right) +
    \mathrm{i}\gamma^{\bar{2}}
    \frac{\partial_{\phi}}{r}+\mathrm{i}\gamma^{\bar{3}}\partial_{z}-m
  \right],
\end{align}
where $V_\mathrm{V,A}=\left(V_{\mathrm{L}}\pm V_{\mathrm{R}}\right)/2$ are
the vector and axial parts of the effective potentials. Note that
the operator $\mathcal{D}$ in eq.~(\ref{eq:Direqgen}) is analogous to that
recently studied in ref.~\cite{Bak12}.

Since eq.~(\ref{eq:Direqgen}) does not explicitly depend on $t$,
$\phi$, and $z$, we shall look for its solution in the form,
\begin{equation}\label{eq:psipsir}
  \psi =
  \exp
  \left(
    -\mathrm{i}Et+\mathrm{i}J_{z}\phi+\mathrm{i}p_{z}z
  \right)
  \psi_{r},
\end{equation}
where $\psi_{r}=\psi_{r}(r)$ is the wave function depending on the
radial coordinate. The values
of $J_{z}$ in eq.~(\ref{eq:psipsir}) were found in ref.~\cite{SchWieGre83} to be $\pm1/2,\pm3/2,\dotsc$.%
%\begin{comment}
%It is convenient to choose $J_{z}=(1/2-l)\text{sgn}(g^{0})$, where
%$l=0,\pm1,\pm2,\dotsc$.
%\end{comment}

It is convenient to rewrite eq.~(\ref{eq:Direqgen}) as
\begin{equation}\label{eq:psir}
  \left[\gamma^{\bar{a}}Q_{a}-m+V\right]\psi_{r}=0,
\end{equation}
where $Q^{a}=q^{a}-q_{\mathrm{eff}}A_{\mathrm{eff}}^{a}$, $q_{\mathrm{eff}}$
is the effective electric charge, $q^{a} = \left( E+J_{z}\omega-V_\mathrm{V}, -\mathrm{i}\partial_{r}, 0, p_{z} \right)$,
\begin{equation}\label{eq:Aeff}
  A_{\mathrm{eff}}^{a} =
  \left(
    0, \frac{\mathrm{i}}{2q_{\mathrm{eff}}r},
    \frac{1}{q_{\mathrm{eff}}}
    \left[
      V_\mathrm{V}\omega r - \frac{J_{z}}{r}
    \right],
    0
  \right)
\end{equation}
is the potential of the effective electromagnetic field, and $V=V_\mathrm{A}\left(\gamma^{\bar{0}}-\omega r\gamma^{\bar{2}}\right)\gamma^{\bar{5}}$.

Let us look for the solution of eq.~(\ref{eq:psir}) in the form,
$\psi_{r}=\left[\gamma^{\bar{a}}Q_{a}+m-V\right]\Phi$, where $\Phi$
is the new spinor. The equation for $\Phi$ reads
\begin{multline}\label{eq:Phieq}
  \bigg[
    \left(
      \partial_{r}+\frac{1}{2r}
    \right)^{2} +
    \left(
      E+J_{z}\omega-V_\mathrm{V}
    \right)^{2} -
    \left(
      \frac{J_{z}}{r}-V_\mathrm{V}\omega r
    \right)^{2} +
    \left(
      V_\mathrm{V}\omega+\frac{J_{z}}{r^{2}}
    \right)
    \Sigma_{3}
    \\
    + 2 V_\mathrm{A} \gamma^{\bar{5}}
    \left[
      \left(
        \frac{J_{z}}{r}-V_\mathrm{V}\omega r
      \right)
      \omega r -
      \left(
        E+J_{z}\omega-V_\mathrm{V}
      \right) +
      \frac{\omega}{2}\Sigma_{z}
    \right]
    - p_{z}^{2}
    \\
    + V_\mathrm{A}^2
    \left(
      1-\omega^{2}r^{2}
    \right)
    + 2mV_\mathrm{A}
    \left(
      \gamma^{\bar{0}}-\omega r\gamma^{\bar{2}}
    \right)
    \gamma^{\bar{5}}-m^{2}
  \bigg]
  \Phi = 0,
\end{multline}
where $\Sigma_{3}=\gamma^{\bar{0}}\gamma^{\bar{3}}\gamma^{\bar{5}}$.

The solution of eq.~(\ref{eq:Phieq}) can be found for ultrarelativistic
particles. In the limit $m\to0$, we can represent $\Phi=v\varphi$
in eq.~(\ref{eq:Phieq}), where $\varphi=\varphi(r)$ is a scalar
function and $v$ is a constant spinor satisfying $\Sigma_{3}v=\sigma v$
and $\gamma^{\bar{5}}v=\chi v$, with $\sigma=\pm1$ and $\chi=\pm1$,
since both $\Sigma_{3}$ and $\gamma^{\bar{5}}$ now commute with
the operator of eq.~(\ref{eq:Phieq}).

Let us first study left particles, $(1+\gamma^{\bar{5}})\psi=0$, corresponding
to $\chi=+1$. The case of right particles with $\chi=-1$ can be
studied analogously. Using the new variable $\rho=|V_{\mathrm{L}}|\omega r^{2}$
we can write the equation for $\varphi_{\sigma}$ as
\begin{equation}\label{eq:phisigma}
  \left[
    \rho\partial_{\rho}^{2}+\partial_{\rho}-\frac{1}{4\rho}
    \left(
      l+\frac{\sigma}{2}\text{sgn}(V_{\mathrm{L}})-\frac{1}{2}
    \right)^{2} -
    \frac{\rho}{4}-\frac{1}{2}
    \left(
      l-\frac{\sigma}{2}\text{sgn}(V_{\mathrm{L}})-\frac{1}{2}
    \right) +
    \kappa
  \right]
  \varphi_{\sigma} = 0,
\end{equation}
where
\begin{align}\label{eq:kappa}
  \kappa = & \frac{1}{4|V_{\mathrm{L}}|\omega}
  \left[
    E^{2}+2E
    \left(
      J_{z}\omega-V_{\mathrm{L}}
    \right) -
    p_{z}^{2}-m^{2} +
    \left(
      V_{\mathrm{L}}
    \right)^{2} +
    J_{z}^{2}\omega^{2}+V_{\mathrm{L}}\omega\sigma
  \right]
  \nonumber
  \\
  & +
  \frac{1}{2}
  \left(
    l-\frac{\sigma}{2}\text{sgn}(V_{\mathrm{L}})-\frac{1}{2}
  \right).
\end{align}
In eq.~(\ref{eq:phisigma}) we choose $J_{z}=(1/2-l)\text{sgn}(V_{\mathrm{L}})$,
where $l=0,\pm1,\pm2,\dotsc$.

Assuming that $\varphi_{\sigma}\to0$ at $r\to\infty$, we get that
$\varphi_{+}=I_{N,s}$ and $\varphi_{-}=I_{N-1,s}$, for $V_{\mathrm{L}}>0$,
as well as $\varphi_{+}=I_{N-1,s}$ and $\varphi_{-}=I_{N,s}$, for
$V_{\mathrm{L}}<0$. Here $N=0,1,2\dotsc$, $s=N-l$, and $I_{N,s}=I_{N,s}(\rho)$
is the Laguerre function%
\footnote{The definition of the Laguerre function is given, e.g., in ref.~\cite{Dvo14}.%
}. The energy spectrum can be found if we take that $\kappa=N$ in
eq.~(\ref{eq:kappa}). We can present $E$ is the form,
\begin{equation}\label{eq:Energy}
 E = V_{\mathrm{L}} +
 \left(
   l-\frac{1}{2}
 \right)
 \omega\text{sgn}(V_{\mathrm{L}})\pm\mathcal{E},
 \quad
 \mathcal{E}=\sqrt{p_{z}^{2}+m^{2}+4|V_{\mathrm{L}}|N\omega}.
\end{equation}
One can see that the neutrino energy depends on the sign of $V_{\mathrm{L}}$.
Note that one should understand $m\neq0$ in eq.~(\ref{eq:Energy})
in the perturbative sense.

The total radial wave function $\psi_{r}$ can be found in the explicit
form if we choose the spinors $v_{\sigma}$ as
\begin{equation}\label{eq:vsigma}
  v_{+}^\mathrm{T} =
  \left(
    1, 0, 0, 0
  \right),
  \quad
  v_{-}^\mathrm{T} =
  \left(
    0, 1, 0, 0
  \right).
\end{equation}
In eq.~(\ref{eq:vsigma}) we assume that the Dirac matrices are in
the chiral representation~\cite{ItzZub80},
\begin{equation}\label{eq:chirrep}
  \gamma^{\bar{0}} =
  \left(
    \begin{array}{cc}
      0 & -1\\
      -1 & 0
    \end{array}
  \right),
  \quad
  \gamma^{\bar{k}} =
  \left(
    \begin{array}{cc}
      0 & \sigma_{k}\\
      -\sigma_{k} & 0
    \end{array}
  \right),
  \quad
  \gamma^{\bar{5}} =
  \left(
    \begin{array}{cc}
      1 & 0\\
      0 & -1
    \end{array}
  \right),
\end{equation}
where $\sigma_{k}$ are the Pauli matrices.

Using eq.~(\ref{eq:vsigma}), we get the radial wave functions
corresponding to different spin projections,
\begin{equation}\label{eq:psireta}
  \psi_{r}^{\pm}=
  \left(
    \begin{array}{c}
      0\\
      \eta_{\pm}
    \end{array}
  \right),
  \quad
  \eta_{+} = \Pi
  \left(
    \begin{array}{c}
      \varphi_{+}\\
      0
    \end{array}
  \right),
  \quad
  \eta_{-} = \Pi
  \left(
    \begin{array}{c}
      0\\
      \varphi_{-}
    \end{array}
  \right),
\end{equation}
where
\begin{equation}\label{eq:Pidef}
  \Pi =
  \left(
    \begin{array}{cc}
      p_{z}-E-J_{z}\omega+V_{\mathrm{L}} & -\mathrm{i}R_{\pm} \\
      -\mathrm{i}R_{\mp} & -p_{z}-E-J_{z}\omega+V_{\mathrm{L}}
    \end{array}
  \right),
\end{equation}
and
\begin{align}
  R_{+}= &
  \partial_{r}+\frac{1}{2r}+\frac{J_{z}}{r}-V_{\mathrm{L}}\omega r =
  \sqrt{|V_{\mathrm{L}}|\omega\rho}
  \left(
    2\partial_{\rho}-\frac{l-1}{\rho}-1
  \right),
  \nonumber
  \\
  R_{-} = &
  \partial_{r}+\frac{1}{2r}-\frac{J_{z}}{r}+V_{\mathrm{L}}\omega r =
  \sqrt{|V_{\mathrm{L}}|\omega\rho}
  \left(
    2\partial_{\rho}+\frac{l}{\rho}+1
  \right).
\end{align}
In eq.~(\ref{eq:Pidef}) the upper signs stay for $V_{\mathrm{L}}>0$
and the lower ones for $V_{\mathrm{L}}<0$.

The properly normalized two component spinors in eq.~\eqref{eq:psireta} have the form,
\begin{equation}\label{eq:etapmposg}
  \eta_{+} = C_{+}
  \left(
    \begin{array}{c}
      \mp
      \left[
        \mathcal{E}\mp p_{z}
      \right]
      I_{N,s}
      \\
      -2\mathrm{i}\sqrt{|V_{\mathrm{L}}|\omega N}I_{N-1,s}
    \end{array}
  \right),
  \quad
  \eta_{-} = C_{-}
  \left(
    \begin{array}{c}
      2\mathrm{i}\sqrt{|V_{\mathrm{L}}|\omega N}I_{N,s}
      \\
      \mp
      \left[
        \mathcal{E}\pm p_{z}
      \right]
      I_{N-1,s}
    \end{array}
  \right),
\end{equation}
for $V_{\mathrm{L}}>0$, and
\begin{equation}\label{eq:etapmnegg}
  \eta_{+} = C_{+}
  \left(
  \begin{array}{c}
    \mp
    \left[
      \mathcal{E}\mp p_{z}
    \right]
    I_{N-1,s}
    \\
    2\mathrm{i}\sqrt{|V_{\mathrm{L}}|\omega N}I_{N,s}
    \end{array}
  \right),
  \quad
  \eta_{-} = C_{-}
  \left(
    \begin{array}{c}
      -2\mathrm{i}\sqrt{|V_{\mathrm{L}}|\omega N}I_{N-1,s}
      \\
      \mp
      \left[
        \mathcal{E}\pm p_{z}
      \right]
      I_{N,s}
    \end{array}
  \right),
\end{equation}
for $V_{\mathrm{L}}<0$. The signs in eqs.~(\ref{eq:etapmposg})
and~(\ref{eq:etapmnegg}) are correlated with the sign in eq.~(\ref{eq:Energy}).

It should be noted that the spinors $\eta_{+}$ and $\eta_{-}$ in
eqs.~(\ref{eq:etapmposg}) and~(\ref{eq:etapmnegg}) are linearly
dependent as it should be for ultrarelativistic particles. Thus, one
can use any independent pair of $\eta_{+}$ and $\eta_{-}$. As usual,
we shall attribute $\eta_{-}$ with the upper sign to a particle degree
of freedom and $\eta_{+}$ with the lower sign to antiparticles. This
choice of independent spinors is convenient for $N>0$. If $N=0$,
one can better use $\eta_{+}$ for $V_{\mathrm{L}}>0$ and $\eta_{-}$
for $V_{\mathrm{L}}<0$ as independent degrees of freedom.

Using the normalization condition for the total wave function,
\begin{equation}\label{eq:normal}
  \int\psi_{N,s,p_{z}}^{\dagger}(x)\psi_{N',s',p'_{z}}(x)\sqrt{-g}\mathrm{d}^{3}x =
  \delta_{NN'}\delta_{ss'}\delta(p_{z}-p'_{z}),
\end{equation}
which includes the dependence on $p_{z}$ and $\phi$, we get the coefficients $C_{\sigma}$
in eqs.~(\ref{eq:etapmposg}) and~(\ref{eq:etapmnegg}) as
\begin{equation}\label{eq:Cpm}
  C_{\sigma}^{2} =
  \frac{|V_{\mathrm{L}}|\omega}{2\pi\mathcal{E}(\mathcal{E}\mp\sigma p_{z})},
\end{equation}
which is valid for any sign of $V_{\mathrm{L}}$.

On the basis of eqs.~(\ref{eq:etapmposg}) and~(\ref{eq:etapmnegg})
one can notice that, at $N=0$, $p_{z}$ is correlated with the particle
helicity. Using eq.~(\ref{eq:Energy}) with $m=0$ as well as eqs.~(\ref{eq:etapmposg})
and~(\ref{eq:etapmnegg}), one can find the possible values of $p_{z}$
at $N=0$. For the convenience, they are listed in table~\ref{tab:g0LRpz}.
Note that at $N>0$, $-\infty<p_{z}<+\infty$.

\begin{table}
  \centering
  \begin{tabular}{|c|c|}
    \hline
    Values of $V_{\mathrm{L,R}}$ & Allowed value of $p_{z}$\tabularnewline
    \hline
    \hline
    \multicolumn{2}{|l|}{Left particles}\tabularnewline
    \hline
    $V_{\mathrm{L}}>0$ & $p_{z}<0$\tabularnewline
    \hline
    $V_{\mathrm{L}}<0$ & $p_{z}>0$\tabularnewline
    \hline
    \multicolumn{2}{|l|}{Left antiparticles}\tabularnewline
    \hline
    $V_{\mathrm{L}}>0$ & $p_{z}>0$\tabularnewline
    \hline
    $V_{\mathrm{L}}<0$ & $p_{z}<0$\tabularnewline
    \hline
    \multicolumn{2}{|l|}{Right particles}\tabularnewline
    \hline
    $V_{\mathrm{R}}>0$ & $p_{z}>0$\tabularnewline
    \hline
    $V_{\mathrm{R}}<0$ & $p_{z}<0$\tabularnewline
    \hline
    \multicolumn{2}{|l|}{Right antiparticles}\tabularnewline
    \hline
    $V_{\mathrm{R}}>0$ & $p_{z}<0$\tabularnewline
    \hline
    $V_{\mathrm{R}}<0$ & $p_{z}>0$\tabularnewline
    \hline
  \end{tabular}
  \protect\caption{Allowed values of $p_{z}$ in the ground state $N=0$ for different
  values of the effective potentials $V_{\mathrm{L,R}}$ for left and
  right particles and antiparticles.\label{tab:g0LRpz}}
\end{table}

It should be noted that $p_{z}$ in table~\ref{tab:g0LRpz} is a
formal quantum number. The physical value of $p_{z}$ for antiparticles
is opposite to that shown in table~\ref{tab:g0LRpz}: $p_{z}^{(\text{phys.})}=-p_{z}$.
Otherwise the electric charge would not be conserved.

Right particles can be treated in the same way as left ones. That
is why we just present only the final results. The expression for
the energy has the same structure as eq.~(\ref{eq:Energy}) with
the replacement $V_{\mathrm{L}}\to V_{\mathrm{R}}$. In eq.~(\ref{eq:psireta})
one has $\psi_{r}^{\pm}=\left(\xi_{\pm},0\right)^{\mathrm{T}}$, where
\begin{equation}\label{eq:xipmposg}
  \xi_{+} = C_{+}
  \left(
    \begin{array}{c}
      \mp
      \left[
        \mathcal{E}\pm p_{z}
      \right]I_{N,s}
      \\
      2\mathrm{i}\sqrt{|V_{\mathrm{R}}|\omega N}I_{N-1,s}
    \end{array}
  \right),
  \quad
  \xi_{-} = C_{-}
  \left(
    \begin{array}{c}
      -2\mathrm{i}\sqrt{|V_{\mathrm{R}}|\omega N}I_{N,s}
      \\
      \mp
      \left[
        \mathcal{E}\mp p_{z}
      \right]
      I_{N-1,s}
    \end{array}
  \right),
\end{equation}
for $V_{\mathrm{R}}>0$, and
\begin{equation}\label{eq:xipmnegg}
  \xi_{+} = C_{+}
  \left(
    \begin{array}{c}
      \mp
      \left[
        \mathcal{E}\pm p_{z}
      \right]
      I_{N-1,s}
      \\
      -2\mathrm{i}\sqrt{|V_{\mathrm{R}}|\omega N}I_{N,s}
    \end{array}
  \right),
  \quad
  \xi_{-} = C_{-}
  \left(
    \begin{array}{c}
      2\mathrm{i}\sqrt{|V_{\mathrm{R}}|\omega N}I_{N-1,s}
      \\
      \mp
      \left[
        \mathcal{E}\mp p_{z}
      \right]
      I_{N,s}
    \end{array}
  \right),
\end{equation}
for $V_{\mathrm{R}}<0$. The argument of the Laguerre functions is
$\rho=|V_{\mathrm{R}}|\omega r^{2}$ now. The new normalization constant
in eq.~(\ref{eq:xipmposg}) and~(\ref{eq:xipmnegg}) reads
\begin{equation}\label{eq:CpmR}
  C_{\sigma}^{2} =
  \frac{|V_{\mathrm{R}}|\omega}{2\pi\mathcal{E}(\mathcal{E}\pm\sigma p_{z})}.
\end{equation}
The signs in eqs.~(\ref{eq:xipmposg})-(\ref{eq:CpmR}) are correlated
with the signs in the expression for the energy.

As in the case of left particles, for right fermions we have that
$-\infty<p_{z}<+\infty$ at $N>0$. The allowed values of $p_{z}$
at $N=0$ are shown in table~\ref{tab:g0LRpz}.

\section{Induced electric current along the rotation axis\label{sec:CURRCALC}}

In this section we show that there is a nonzero induced electric current
flowing along the rotation axis in the system of electroweakly interacting
particles. In our calculation we shall use the exact solution of the
Dirac equation obtained in section~\ref{sec:DIRACEQ}.

In section~\ref{sec:DIRACEQ} we already mentioned that there is
a correlation between $p_{z}$ and the helicity at $N=0$. Thus one
expects that there can be macroscopic fluxes of particles in the rotating
matter. Let us first examine this issue for left fermions. We shall
calculate the mean hydrodynamic currents of particles and antiparticles
with respect to the coordinates $x^{\mu}$ in the rotating frame.
These currents have the form,
\begin{equation}\label{eq:hydcurrgen}
  j_{\mathrm{L}f,\bar{f}}^{\mu} =
  \sum_{N,s=0}^{\infty}
  \int\mathrm{d}p_{z}
  \bar{\psi}\gamma^{\mu}(x)\psi
  \rho_{f,\bar{f}}^\mathrm{L}(\mathcal{E}\pm V_{\mathrm{L}}),
\end{equation}
where $\rho_{f,\bar{f}}^\mathrm{L}(E) = \left\{ \exp(\beta[E\mp\mu_{\mathrm{L}}])+1 \right\}^{-1}$
is the Fermi-Dirac distribution for fermions, with the lower sign staying for antifermions, $\beta=1/T$
is the reciprocal temperature of the fermion gas, and $\mu_{\mathrm{L}}$
is the chemical potential of left particles. The spinor in eq.~(\ref{eq:hydcurrgen})
corresponds to the exact solution of the Dirac equation in eqs.~(\ref{eq:etapmposg})
and~(\ref{eq:etapmnegg}). Note that, for the first time, this method
for the calculation of the current was proposed in ref.~\cite{Vil80}.

We will be interested in the expression for $j_{\mathrm{L}f}^{\mu}$
linear in $\omega$. That is why we use $\mathcal{E}+V_{\mathrm{L}}$
instead of the total particle energy, cf. eq.~(\ref{eq:Energy}),
in the distribution function. The contribution of the noninertial part of the energy $\sim \omega \text{sgn}(V_\mathrm{L,R})(l - 1/2)$, see eq.~\eqref{eq:Energy}, to the currents is computed in appendix~\ref{sec:CURRDER}. We should study only $j_{\mathrm{L}f}^{3}$
since it is this component of the current that is linear in $\omega$.

Using the orthogonality of Laguerre functions,
\begin{equation}
  \sum_{s=0}^{\infty}I_{N,s}(\rho)I_{N',s}(\rho)=\delta_{NN'},
\end{equation}
as well as eqs.~(\ref{eq:vierbein}), (\ref{eq:psipsir}), (\ref{eq:chirrep}),
(\ref{eq:psireta}), (\ref{eq:etapmposg}), (\ref{eq:etapmnegg}),
and~(\ref{eq:Cpm}) we get that
\begin{equation}\label{eq:sums}
  \sum_{s=0}^{\infty}
  \bar{\psi}\gamma^{3}(x)\psi =
  \pm\frac{p_{z}}{\mathcal{E}}\frac{|V_{\mathrm{L}}|\omega}{\pi},
\end{equation}
where the upper sign stays for particles and the lower one for antiparticles.

On the basis of eq.~(\ref{eq:sums}) and table~\ref{tab:g0LRpz}
we find that only the lowest level $N=0$ contributes to the current.
Finally we get that
\begin{equation}\label{eq:J3lL}
  j_{\mathrm{L}f}^{3} =
  -\frac{V_{\mathrm{L}}\omega}{\pi}
  \int_{0}^{\infty}\mathrm{d}p
  \rho_{f}^\mathrm{L}(p+V_{\mathrm{L}}).
\end{equation}
Analogously to eq.~(\ref{eq:J3lL}) we can obtain the following expression:
\begin{equation}\label{eq:J3antilL}
 j_{\mathrm{L}\bar{f}}^{3} =
 -\frac{V_{\mathrm{L}}\omega}{\pi}
 \int_{0}^{\infty}\mathrm{d}p
 \rho_{\bar{f}}^\mathrm{L}(p-V_{\mathrm{L}}),
\end{equation}
which is valid for antifermions.

The contribution to the hydrodynamic current from right femions is
analogous to eqs.~(\ref{eq:J3lL}) and~(\ref{eq:J3antilL}). It
is
\begin{equation}\label{eq:J3lR}
  j_{\mathrm{R}f}^{3} =
  \frac{V_{\mathrm{R}}\omega}{\pi}
  \int_{0}^{\infty}\mathrm{d}p
  \rho_{f}^\mathrm{R}(p+V_{\mathrm{R}}),
\end{equation}
for particles, and
\begin{equation}\label{eq:J3antilR}
  j_{\mathrm{R}\bar{f}}^{3} =
  \frac{V_{\mathrm{R}}\omega}{\pi}
  \int_{0}^{\infty}\mathrm{d}p
  \rho_{\bar{f}}^\mathrm{R}(p-V_{\mathrm{R}}),
\end{equation}
for antiparticles. Here $\rho_{f,\bar{f}}^\mathrm{R}(E)$ are the distributions of right particles and antiparticles which can be obtained from $\rho_{f,\bar{f}}^\mathrm{L}(E)$ by replacing $\mu_{\mathrm{L}} \to \mu_{\mathrm{R}}$, where $\mu_{\mathrm{R}}$ is the chemical potential
of right fermions.
%Note that eqs.~(\ref{eq:J3lL})-(\ref{eq:J3antilR}) are valid for any temperature.

Now, using eqs.~(\ref{eq:J3lL})-(\ref{eq:J3antilR}), we can obtain
the expression for the third component of the electric current as
$J^{3} = q_{f} \left( j_{\mathrm{L}f}^{3} - j_{\mathrm{L}\bar{f}}^{3} + j_{\mathrm{R}f}^{3} - j_{\mathrm{R}\bar{f}}^{3} \right)$,
where $q_{f}$ is the electric charge of the fermion $f$ including
the sign. For example, $q_{e}=-e$ for an electron, $q_{u}=2e/3$
for an $u$ quark, and $q_{d}=-e/3$ for a $d$ quark. Here $e>0$
is the absolute value of the elementary electric charge. In the expression
for $J^{3}$, we use the convention that the direction of the electric
current coincides with the motion of the positive electric charge. Finally we get for the electric current,
\begin{equation}\label{eq:elcurr}
  \mathbf{J} =
  \frac{q_{f}\bm{\omega}}{\pi}
  \left(
    V_{\mathrm{R}}\mu_{\mathrm{R}}-V_{\mathrm{L}}\mu_{\mathrm{L}}
  \right),
\end{equation}
where we restore vector notations. We remind that in eq.~(\ref{eq:elcurr})
we keep only the terms linear in $\omega$ and $V_{\mathrm{L,R}}$. Some details of the derivation of eq.~\eqref{eq:elcurr} from eqs.~(\ref{eq:J3lL})-(\ref{eq:J3antilR}) are provided in appendix~\ref{sec:CURRDER}.

We can attribute the existence of the induced electric current in
rotating matter, where the parity violating interaction is present,
to the new GRE; cf. section~\ref{sec:INTR}.

\section{Generation of TMF and magnetic helicity in a pulsar\label{sec:TMFGENERATION}}

In this section we apply GRE for the calculation of TMF inside a
pulsar. We also briefly consider the generation of the magnetic helicity
in a compact rotating star.

If we consider a rapidly rotating compact astrophysical object, like
a neutron star (NS) or even a hypothetical quark star (QS), then the
mechanism described in section~\ref{sec:CURRCALC} will induce the
electric current along the rotation axis of such a star. We shall
suppose that this current forms a closed circuit connected somewhere at the stellar surface.
Thus, using the Maxwell equation $(\nabla\times\mathbf{B})=\mathbf{J}$,
we get that this current should induce a TMF, $B_{\mathrm{tor}}\sim RJ$,
where $R\sim10\thinspace\text{km}$ is the stellar radius.

It should be noted that a compact star typically has a poloidal magnetic
field $B_{\mathrm{pol}}$, which is measured in astronomical observations.
For instance, the radiation of a pulsar can be explained by the emission
of electromagnetic waves by the rotating magnetic dipole associated
with $B_{\mathrm{pol}}$, provided there is a nonzero angle between
$\mathbf{B}_{\mathrm{pol}}$ and $\bm{\omega}$. However, as shown
in ref.~\cite{LanJon09}, using general arguments for the magnetohydrodynamic
equilibrium of an axisymmetric NS, a purely poloidal magnetic field
configuration turns out to be unstable. Thus an internal nonzero TMF
should exist in a compact star.

Let us first consider the generation of TMF in NS composed of degenerate
electrons and nucleons, like neutrons and protons. In this case $u$
and $d$ quarks are confined inside nucleons. The typical electron
density in NS is $n_{e}\approx9\times10^{36}\thinspace\text{cm}^{-3}$,
which corresponds to the electron fraction $Y_{e}\approx0.05$. It
gives the chemical potential of electrons $\mu_{e}\approx125\thinspace\text{MeV}\gg m_{e}$.
Therefore electrons are ultrarelativistic whereas neutrons and protons
are nonrelativistic. Note that the nonzero electron mass was shown in ref.~\cite{ChaZhi10} to slightly contribute to the induced electric
current. Thus we can assume that electrons
are approximately massless in NS and the results of sections~\ref{sec:DIRACEQ}
and~\ref{sec:CURRCALC} are valid.

Since the chiral symmetry is unbroken, we can consider left and right
chiral projections as independent degrees of freedom. For simplicity
we shall take that left and right electrons are in equilibrium with
$\mu_{\mathrm{L}}\sim\mu_{\mathrm{R}}\sim\mu_{e}$. The situations,
when the chiral imbalance $\mu_5=(\mu_{\mathrm{R}}-\mu_{\mathrm{L}})/2 \neq 0$
is important in NS, are studied, e.g., in refs.~\cite{DvoSem15,ChaZhi10}.
Since $n_{p}\ll n_{n}$ in NS, $(V_{\mathrm{L}}-V_{\mathrm{R}}) = G_{\mathrm{F}} n_{n} / \sqrt{2} \approx12\thinspace\text{eV}$
for $n_{n}\approx1.8\times10^{38}\thinspace\text{cm}^{-3}$. Eventually,
taking that $\omega\sim10^{3}\thinspace\text{s}^{-1}$ and $R\sim10\thinspace\text{km}$
as well as using eq.~(\ref{eq:elcurr}), we get that $B_{\mathrm{tor}}\approx2.5\times10^{8}\thinspace\text{G}$
can be generated in a rotating NS.

The obtained value of $B_{\mathrm{tor}}$ is comparable with $B_{\mathrm{pol}} \lesssim (10^{8} - 10^{9}) \thinspace\text{G}$
in weakly magnetized old millisecond pulsars~\cite{PhiKul94}. It
should be noted that the stability of magnetic fields in NS can be
reached if $0\leq E_{\mathrm{tor}}/E_{\mathrm{mag}}<0.07$~\cite{LanJon09},
where $E_{\mathrm{tor}}\sim B_{\mathrm{tor}}^{2}$ is the energy of
TMF and $E_{\mathrm{mag}}\sim(B_{\mathrm{tor}}^{2}+B_{\mathrm{pol}}^{2})$
is the total magnetic energy. Our estimate for $B_{\mathrm{tor}}$
satisfies this criterion.

Let us discuss the creation of TMF in a hypothetical QS. Although
QSs have not been observed yet, their properties are actively studied
theoretically~\cite{Web05}. Various models of QS predict that it
consists of free $u$ and $d$ quarks with some admixture of $s$
quarks. After the analysis of
various equations of state of QS matter, the strangeness fraction
was found in ref.~\cite{MenProMen06} not to exceed $\sim0.3$. Thus we can approximately omit the contribution
of $s$ quarks in the calculation of the induced current.

We shall suppose that inside QS we have $n_{u}=n_{0}/3\approx0.6\times10^{38}\thinspace\text{cm}^{-3}$
and $n_{d}=2n_{0}/3\approx1.2\times10^{38}\thinspace\text{cm}^{-3}$,
where $n_{0}\approx1.8\times10^{38}\thinspace\text{cm}^{-3}$ is the
nuclear density. At such high densities the chiral symmetry can be
unbroken in QS~\cite{NgCheChu03}, allowing one to consider the independent chiral projections of the wave functions of $u$ and
$d$ quarks. Therefore we can again use the
results of sections~\ref{sec:DIRACEQ} and~\ref{sec:CURRCALC}.
As in case of NS, we can also assume that left and right quarks are
in equilibrium, $\mu_{u,d}^{\mathrm{L}}\sim\mu_{u,d}^{\mathrm{R}}\sim\mu_{u,d}$,
just for simplicity. Here $\mu_{u}=(1/3)^{1/3}\mu_{0}\approx 235\thinspace\text{MeV}$
and $\mu_{d}=(2/3)^{1/3}\mu_{0}\approx 296\thinspace\text{MeV}$, where
$\mu_{0} = 339\thinspace\text{MeV}$.

Finally, using, in eq.~(\ref{eq:elcurr}), the adopted values of
densities and chemical potentials of quarks, the values of $V_{\mathrm{L,R}}$
in table~\ref{tab:VLR} as well as for $\omega\sim10^{3}\thinspace\text{s}^{-1}$
and $R\sim10\thinspace\text{km}$, we get that $B_{\mathrm{tor}} \approx 4.9 \times 10^{8} \thinspace \text{G}$
can exist inside a rotating QS. The obtained value of TMF is slightly
greater than $B_{\mathrm{tor}}$ for NS. Note that the derived strength
of TMF is also in agreement with stability criterion obtained in ref.~\cite{LanJon09}
for old weakly magnetized millisecond pulsars~\cite{PhiKul94}.

It should be noted that, in our estimate of TMF in QS, we account
for only $(ud)$ and $(du)$ contributions to the electric current.
Using the analogy of the rotating electroweak matter with the presence
of an effective magnetic field~\cite{GriSavStu07} (see also eqs.~(\ref{eq:psir}) and~\eqref{eq:Aeff} in section~\ref{sec:DIRACEQ})
and the results of ref.~\cite{Dvo14b}, we get that $(uu)$ and $(dd)$
interactions do not contribute to the current in eq.~(\ref{eq:elcurr}).

We should mention that, in the generation of TMF in a compact star,
we discuss a thermally relaxed stage in the evolution of this astrophysical
object. This approximation is valid since we consider old millisecond
pulsars with ages $\sim (10^8 - 10^{10})\thinspace\text{yr}$~\cite{PhiKul94}.
It means that we discard any possible effects related to turbulence
which should be treated on the basis of the Navier-Stokes equation.
In our analysis we also do not consider a differential rotation either.

The creation of TMF in a compact star is closely related to the problem
of the generation of the magnetic helicity defined as
\begin{equation}\label{eq:heldef}
  H = \int\mathrm{d}^{3}x
  \left(
    \mathbf{A}\cdot\mathbf{B}
  \right),
\end{equation}
where $\mathbf{A}$ is the 3D vector potential. If there is configuration
of magnetic fields in a star consisting of toroidal and poloidal fields,
then $H$ in eq.~(\ref{eq:heldef}) has the form~\cite{Ber99},
$H=2L\Phi_{\mathrm{tor}}\Phi_{\mathrm{pol}}$, where $\Phi_{\mathrm{tor}}$
and $\Phi_{\mathrm{pol}}$ are the fluxes of toroidal and poloidal
fields and $|L|=1$ is the linkage number. The magnetic helicity is
a conserved quantity in a perfectly conducting medium. It is this
fact which provides the stability of a poloidal field in a compact
star. Note that another mechanism for the generation of the magnetic
helicity in a nonrotating NS, based on the electron-nucleon electroweak
interaction, was recently proposed in refs.~\cite{DvoSem15,DvoSem15JCAP}.

\section{Pulsar kicks due to the asymmetric neutrino emission\label{sec:KICKS}}

In this section we shall use GRE for the description of the asymmetry in the neutrino emission from NS. We shall also consider the applicability of our results to explain linear velocities of pulsars.

It is well established that some pulsars have great linear velocities up to $\sim 10^3 \thinspace\text{km} \cdot \text{s}^{-1}$~\cite{Hob05}. There are various models for the explanation of this phenomenon based on, e.g., the asymmetric electromagnetic radiation~\cite{HarTad75} and the asymmetric explosion leading to the anisotropic neutrino emission~\cite{Bur07}. We also mention ref.~\cite{KusSeg96}, where the asymmetry in neutrino oscillations in matter and an external magnetic field was used to account for pulsar kicks. The idea that anisotropically emitted electrons, owing to CME, pass the momentum to NS, was discussed in ref.~\cite{ChaZhi10}. Nevertheless the origin of peculiar velocities of pulsars is still unclear.

In ref.~\cite{Joh05} it was established that linear velocities of pulsars are correlated with their angular velocities. Thus we may try to apply GRE, which predicts particle fluxes along the rotation axis, to explain pulsar kicks. However, unlike ref.~\cite{ChaZhi10}, we shall examine the possibility of asymmetric neutrino emission since it is not very clear how charged particles can escape NS.

Using eqs.~\eqref{eq:J3lL} and~\eqref{eq:J3antilL}, we get the total hydrodynamic current of neutrinos along the rotation axis as
\begin{equation}\label{eq:j3nu}
  j^3_\mathrm{L} =
  G_\mathrm{F} n_n \frac{\omega T}{\pi\sqrt{2}} F(y),
  \quad
  F(y) =
  \int_0^\infty
  \left(
    \frac{1}{e^{x-y}+1} + \frac{1}{e^{x+y}+1}
  \right)
  \mathrm{d}x =
  2\ln
  \left(
    1 + e^y
  \right) - y,
\end{equation}
where $y = (\mu_\mathrm{L} - V_\mathrm{L})/T$ and $T$ is the neutrino temperature. To derive eq.~\eqref{eq:j3nu} we take into account that $V_\mathrm{L} \approx - G_\mathrm{F} n_n / \sqrt{2}$ for $\nu_e$ (see table~\ref{tab:VLR}) and $n_n \gg n_e$. Note that both $\mu_\mathrm{R}$ and $V_\mathrm{R}$ are equal to zero for ultrarelativistic neutrinos. That is why we account for the contribution of only left neutrinos in eq.~\eqref{eq:j3nu}.

\begin{figure}
  \centering
  \includegraphics[scale=.5]{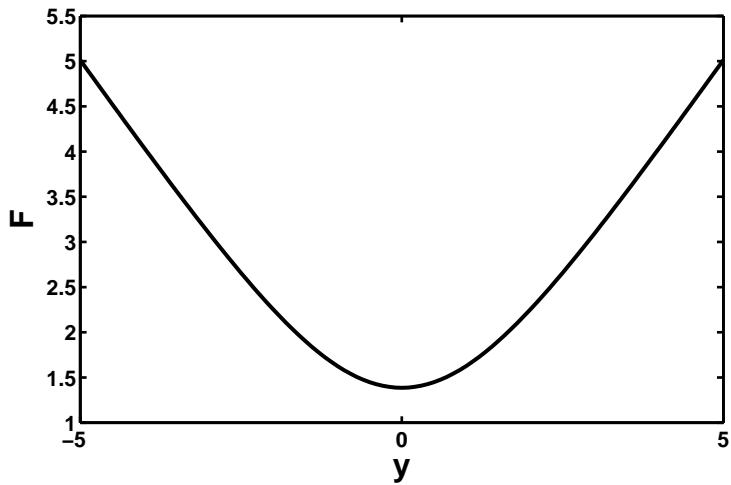}
    \caption{The function $F$ in eq.~\eqref{eq:j3nu} versus $y$.
  \label{fig:F}}
\end{figure}

The more complicated structure of the hydrodynamic current in eq.~\eqref{eq:j3nu} compared to that of the electric current in eq.~\eqref{eq:elcurr} is owing to the fact that $j^3_{\mathrm{L}\bar{f}}$ has the same direction as $j^3_{\mathrm{L}f}$, whereas $J^3_{\mathrm{L}\bar{f}}$ is directed oppositely to $J^3_{\mathrm{L}f}$.
The function $F(y)$ is plotted in figure~\ref{fig:F}. It is interesting to mention that $F(0) \approx 1.4$. Thus, there is a nonzero neutrino flux even at $\mu_\mathrm{L} = 0$.

The simulations carried out in ref.~\cite{Tot98} show that there is a significant nonzero neutrino asymmetry $\Delta n_{\nu_e} = n_{\nu_e} - n_{\bar{\nu}_e}$, owing to direct Urca processes, which lasts up to $t_1 \sim 0.1\thinspace\text{s}$ after the onset of the supernova collapse. When $0<t<t_1$, the typical neutrino energy $E_\nu \sim 10\thinspace\text{MeV}$ and chemical potential of $\nu_e$ is $\mu_\mathrm{L} \sim 10\thinspace\text{MeV}$~\cite{KeiRafJan03}. At $t_1 < t \lesssim t_2 \sim 10^2 \thinspace\text{yr}$ other neutrino species start to be emitted resulting in the diminishing of $\mu_\mathrm{L}$. When $t_2 \lesssim t \lesssim t_3 \sim 10^6 \thinspace\text{yr}$ only $\nu\bar{\nu}$ pairs can be emitted in modified Urca processes~\cite{Yak11} leading to $\mu_\mathrm{L} = 0$. For simplicity we shall assume that $T \sim 10^8 \thinspace \text{K} = \text{const}$ and $E_\nu \sim 10\thinspace\text{keV}$~\cite{Yak11} at this stage of the NS evolution.

The total momentum carried away by neutrinos during the time interval $\Delta t$ is $P \sim j^3 E_\nu S \Delta t$, where $S = \pi R^2$ is area of the equatorial cross section of NS and and $R \sim 10\thinspace\text{km}$ is the NS radius. Using eq.~\eqref{eq:j3nu}, one gets that the greatest $P$ is achieved at $t_2 \lesssim t \lesssim t_3$. As a result, NS will get a recoil velocity $v = P/M$, where $M$ is the NS mass. We shall take $M = 1.44 M_\odot$, where $M_\odot \approx 2 \times 10^{33} \thinspace \text{g}$ is the solar mass, to get the upper bound for $v$. Taking that $\omega \sim 10^3 \thinspace\text{s}^{-1}$, we get that $v \sim 10^{-16}\thinspace\text{cm} \cdot \text{s}^{-1}$. The obtained value is sure to be beyond the possibility of modern astronomical observations. Therefore, despite there is a nonzero anisotropy in the neutrino emission in NS owing to GRE, this effect is unlikely to result in any testable phenomena.

\section{Discussion\label{sec:CONCL}}

In conclusion we note that in the present work we have studied the
evolution of particles, involved in the parity violating electroweak
interaction, in the rotating matter. In section~\ref{sec:DIRACEQ},
we have obtained the new exact solution of the Dirac equation
for a test ultrarelativistic particle, which account for the noninertial effects.
Then, in section~\ref{sec:CURRCALC}, we have computed the induced
electric current along the rotation axis on the basis of the exact
solution of the Dirac equation. In section~\ref{sec:TMFGENERATION},
we have applied our results for the generation of TMF and the magnetic
helicity in compact rotating stars. Finally, in section~\ref{sec:KICKS}, we have considered the production of the anisotropy in the neutrino emission from NS, owing to GRE, and examined the applicability of this effect for the explanation of peculiar velocities of pulsars.

Several new results have been obtained in this work. Firstly, we mention
that the vierbein vectors in eq.~(\ref{eq:vierbein}) have never
been used previously in the Dirac eq.~(\ref{eq:Depsicurv}), which
accounts for the electroweak interaction with background matter in
curved space-time. Another veirbein was recently used in ref.~\cite{Dvo14}.
However, the choice of the vierbein in the present work is likely
to be more appropriate for ultrarelativistic particles in a rotating
frame. In particular, here we have obtained the correct form of the
``centrifugal'' energy, or the energy of the rotation---angular momentum
coupling, $E_{\mathrm{cf}}=-(\mathbf{J}\cdot\bm{\omega})$; cf. eq.~(\ref{eq:Energy}).
The obtained expression for $E_{\mathrm{cf}}$ coincides with the
result of ref.~\cite{HehNi90} derived on the basis of the general
analysis. The form of $E_{\mathrm{cf}}$ obtained in ref.~\cite{Dvo14},
where another vierbein was used, is slightly different. Therefore,
the vierbein adopted in ref.~\cite{Dvo14} is likely to be more appropriate
for the description of nonrelativistic particles in a rotating frame;
cf. ref.~\cite{BakFur10}.

Secondly, we have predicted the new GRE. This effect consists in the
appearance of the electric current in the rotating matter composed
of massless particles involved in the parity violating electroweak
interaction. This electric current flows along the rotation axis.
The new GRE is analogous to CME, known in QED,
which consists in the generation of the electric current of massless
charged particles along the external magnetic field~\cite{Kha15,Vil80}.
It should be noted that, for the first time, the analogy between the motion
in a rotating electroweak matter and in an external magnetic field
was mentioned in ref.~\cite{GriSavStu07}.

Note that the appearance of the electric tension in a rotating conductor
owing to the noninertial effects was also discussed in ref.~\cite{AhmErm02}.
The electric tension, predicted in ref.~\cite{AhmErm02}, is induced
mainly by the Coriolis force acting on charged particles in a rotating
conductor. If $\bm{\omega}$ is chosen along the $z$-axis, this tension
is found in ref.~\cite{AhmErm02} to be along the azimuthal direction.
In our case, the electric current is owing to both the matter rotation
and the presence of the parity violating electroweak interaction.
We predict that the induced electric current flows along the rotation
axis.

We have used the solution of a Dirac equation in the rotating frame
for the calculation of the induced electric current. It means that
this current flows inside the rotating matter since the quantum states
of charged particles are measured by a corotating observer. For example,
if one used the wave functions obtained in ref.~\cite{StuTok14},
although they look similar to those in eqs.~(\ref{eq:etapmposg}) and~(\ref{eq:etapmnegg}),
we would get the electric current with respect the a nonrotating observer,
which cannot be applied for the generation of the internal TMF.

We have used the calculated electric current to generate
TMF and the magnetic helicity inside a rotating compact star, like
NS or QS. The strength of TMF generated turned out to be moderate,
$B_{\mathrm{tor}}\gtrsim10^{8}\thinspace\text{G}$, for both NS and
QS. However, such TMF is comparable with a poloidal field in weakly
magnetized old millisecond pulsars~\cite{PhiKul94}. Note that the
obtained strength of TMF is in agreement with a criterion for the
magnetic field stability derived in ref.~\cite{LanJon09}. It should
be noted that our model for the generation of TMF does not require
the existence of a significant chiral imbalance between left and right
charged particles. Such an imbalance is essential if CME is used to create TMF; cf. ref.~\cite{ChaZhi10}.

Finally, in section~\ref{sec:KICKS}, we analyzed the possibility to apply GRE to explain linear velocities of pulsars by the asymmetric neutrino emission from a rotating NS. We have estimated the total neutrino flux along the rotation axis as well as the recoil velocity of NS. It turned out that a pulsar kick caused by GRE is outside the observationally tested region.

At the end of this section we should make a comment on the influence of nonzero masses of charged particles on the generation of an electric current in a rotating star. It was mentioned in ref.~\cite{ChaZhi10} that the value of the current, induced by CME, slightly diminishes if a small, compared to the energy, but nonzero electron mass is accounted for. A more detailed analysis of the influence on the current, induced in frames of CME, from nonzero masses of flavored fermions, in case of strong interactions, was made in ref.~\cite{HoyNisBan11}. It was shown in ref.~\cite{HoyNisBan11} that CME disappears only in the great masses limit. The effect of the nonzero electron mass on the generation of strong magnetic fields in magnetars was also studied in ref.~\cite{GraKapRed14}.

We should mention that, in case of GRE, a nonzero current in eq.~\eqref{eq:elcurr} exists even at zero chiral imbalance: $\mu_5 = 0$. Therefore, even if an initial $\mu_5(0) \neq 0$ is washed out owing to spin-flip processes taking place at a nonzero mass, as predicted in refs.~\cite{DvoSem15,DvoSem15JCAP,GraKapRed14}, the current will be nonzero due to $V_5 = (V_\mathrm{L} - V_\mathrm{R})/2 \neq 0$. For example, in section~\ref{sec:TMFGENERATION}, we assumed that $\mu_5 = 0$ to simplify the estimates. In the model for the generation of strong magnetic fields in magnetars, elaborated in refs.~\cite{DvoSem15,DvoSem15JCAP}, $\mu_5 \sim - V_5$ appears inside NS in the course of the magnetic fields evolution. In this case, our estimates for $B_\mathrm{tor}$ obtained in section~\ref{sec:TMFGENERATION} will slightly change since $\mu_\mathrm{R} \neq \mu_\mathrm{L}$.

GRE at $V_5 \neq 0$ and $\mu_5 = 0$ is likely to exist in all particular cases we considered in the present work since we discussed ultrarelativistic particles having $m_f / \langle E_f \rangle \ll 1$, where $m_f$ is the fermion mass and $\langle E_f \rangle$ is the typical fermion energy. Nevertheless, a more detailed quantum field theory analysis of this fact, like in ref.~\cite{HoyNisBan11}, is required.

\acknowledgments{
I am thankful to V.G.~Bagrov and V.B.~Semikoz for useful discussions, to A.I.~Studenikin
for communications, to FAPESP (Brazil) for the Grant No.~2011/50309-2,
to the Tomsk State University Competitiveness Improvement Program and to RFBR (research project No.~15-02-00293) for partial support.}

\appendix

\section{Details of the electric current calculation\label{sec:CURRDER}}

In this appendix we derive the electric current in eq.~\eqref{eq:elcurr} and discuss the approximations made.

Let us first consider left fermions. Using eq.~\eqref{eq:hydcurrgen}, one gets that the contribution of particles and antiparticles to the electric current along the rotation axis is
\begin{equation}\label{eq:curr3}
  J_{\mathrm{L}}^{3} =
  q_f
  \sum_{N,s=0}^{\infty}
  \int\mathrm{d}p_{z}
  \left[
    \bar{\psi}_f\gamma^{\mu}(x)\psi_f
    \rho_{f}^\mathrm{L}(E_f) -
    \bar{\psi}_{\bar{f}}\gamma^{\mu}(x)\psi_{\bar{f}}
    \rho_{\bar{f}}^\mathrm{L}(E_{\bar{f}})
  \right],
\end{equation}
where $E_{f,\bar{f}} = \mathcal{E} \pm V_\mathrm{L} \pm \omega \text{sgn}(V_\mathrm{L})(l-1/2)$ are the energies of particles and antiparticles, which also include $E_\mathrm{cf} = - (\mathbf{J}\cdot\bm{\omega})$, and $\psi_{f,\bar{f}}$ are the wave functions of particles and antiparticles.

Assuming that $\mathcal{E} \gg |E_\mathrm{cf}|$, we can expand the distribution function in eq.~\eqref{eq:curr3} in a series
\begin{equation}\label{eq:decdistr}
  \rho_{f}^\mathrm{L}(E_f) =
  \rho_{f}^\mathrm{L}(E_0) -
  \frac{\omega \text{sgn}(V_\mathrm{L}) l'}{T} \frac{\exp(E_0)}{(\exp(E_0)+1)^2} + \dotsc,
\end{equation}
where $E_0 = \mathcal{E} \pm V_\mathrm{L} \mp \omega \text{sgn}(V_\mathrm{L})/2$ and $l' = l$ for the upper sign in $E_0$ or $l'=l-1$ for the lower sign in $E_0$. One can write down the decomposition of $\rho_{\bar{f}}^\mathrm{L}(E_{\bar{f}})$ analogous to that in eq.~\eqref{eq:decdistr}.

Using the explicit form of the wave functions in eqs.~\eqref{eq:etapmposg} and~\eqref{eq:etapmnegg} and the known sum involving the Laguerre function~\cite{SokTer74},
\begin{equation}%\label{eq:decdistr}
  \sum_{s=0}^\infty
  (s-N) I^2_{N,s}(\rho) = \rho,
\end{equation}
we get that accounting for $E_\mathrm{cf}$ is equivalent to the replacement $\mu_\mathrm{L} \to \mu'_\mathrm{L} = \mu_\mathrm{L} + (\omega r)^2 V_\mathrm{L}$. Here we neglected terms $\sim \omega^3$. In fact, the terms $\sim (\omega r)^2$ are the noninertial contributions to the electric current.

Thus the ground state $N=0$ contribution to $J_{\mathrm{L}}^{3}$ in eq.~\eqref{eq:curr3} takes the form,
\begin{equation}\label{eq:curr3grs}
  J_{\mathrm{L}}^{3} =
  - q_f \frac{\omega V_\mathrm{L}}{\pi}
  \int\mathrm{d}p
  \left[
    \rho_{f}^\mathrm{L}(p+V_\mathrm{L}) - \rho_{\bar{f}}^\mathrm{L}(p-V_\mathrm{L})
  \right],
\end{equation}
where we should use $\mu'_\mathrm{L}$ instead of $\mu_\mathrm{L}$. Introducing the new variables $x = p/T$ and $y = (\mu_\mathrm{L} - V_\mathrm{L})/T$ in eq.~\eqref{eq:curr3grs} and using the identity
\begin{equation}%\label{eq:curr3grs}
  \int_0^\infty
  \left(
    \frac{1}{e^{x-y}+1} - \frac{1}{e^{x+y}+1}
  \right)
  \mathrm{d}x = y,
\end{equation}
and eq.~\eqref{eq:curr3grs}, we get that $J_{\mathrm{L}}^{3} = -(q_f/\pi) \omega V_\mathrm{L} (\mu'_\mathrm{L} - V_\mathrm{L})$. The contribution of right fermions can be obtained analogously. It is interesting to mention that the electric current obtained does not depend on the plasma temperature.

In section~\ref{sec:TMFGENERATION} we use the following values of the parameters: $\mu_\mathrm{L,R} \sim 10^2 \thinspace \text{MeV}$, $V_\mathrm{L,R} \sim 10 \thinspace \text{eV}$, $\omega \sim 10^{3} \thinspace \text{s}^{-1}$, and $r < R \sim 10 \thinspace \text{km}$. Therefore, one gets that $\mu_\mathrm{L,R} \gg V_\mathrm{L,R} \gg (\omega r)^2 V_\mathrm{L,R}$. Taking into account these estimates we arrive to eq.~\eqref{eq:elcurr}.


\begin{thebibliography}{100}

\bibitem{Ryd09}
  L.~Ryder,
  \textit{Introduction to General Relativity},
  Cambridge University Press, Cambridge U.K. (2009), 
  pp.~18--46.

\bibitem{Kha15}
  V.A.~Miransky and I.A.~Shovkovy,
  \textit{Quantum field theory in a magnetic field: From quantum chromodynamics to graphene and Dirac semimetals},
  \textit{Phys. Rep.}
  \textbf{576} (2015) 1,
  [arXiv:1503.00732].

\bibitem{Vil80}
  A.~Vilenkin,
  \textit{Equilibrium parity-violating current in a magnetic field},
  \textit{Phys. Rev.}
  \textbf{D 22} (1980) 3080.

\bibitem{DvoSem15}
  M.~Dvornikov and V.B.~Semikoz,
  \textit{Magnetic field instability in a neutron star driven by the electroweak electron-nucleon interaction versus the chiral magnetic effect},
  \textit{Phys. Rev.}
  \textbf{D 91} (2015) 061301
  [arXiv:1410.6676].

\bibitem{BraSub05}
  A.~Brandenburg and K.~Subramanian,
  \textit{Astrophysical magnetic fields and nonlinear dynamo theory},
  \textit{Phys. Rept.}
  \textbf{417} (2005) 1
  [astro-ph/0405052].

\bibitem{FloRud77}
  E.~Flowers and M.A.~Ruderman,
  \textit{Evolution of pulsar magnetic fields},
  \textit{Astrophys. J.}
  \textbf{215} (1977) 302.

\bibitem{Tay73}
  R.J.~Tayler,
  \textit{The adiabatic stability of stars containing magnetic fields-I.Toroidal fields}, \textit{Mon. Not. R. Astron. Soc.}
  \textbf{161} (1973) 365.

\bibitem{Hob05}
  G.~Hobbs, D.R.~Lorimer, A.G.~Lyne and M.~Kramer,
  \textit{A statistical study of 233 pulsar proper motions},
  \textit{Mon. Not. R. Astron. Soc.}
  \textbf{360} (2005) 974
  [astro-ph/0504584].

\bibitem{Joh05}
  S.~Johnston, G.~Hobbs, S.~Vigeland, M.~Kramer, J.M.~Weisberg and A.G.~Lyne,
  \textit{Evidence for alignment of the rotation and velocity vectors in pulsars},
  \textit{Mon. Not. Roy. Astron. Soc.}
  \textbf{364} (2005) 1397
  [astro-ph/0510260].

\bibitem{MohPal04}
  R.N.~Mohapatra and P.B.~Pal,
  \textit{Massive Neutrinos in Physics and Astrophysics},
  World Scientific, Singapore (2004),
  3rd. edn., pp.~5--8.

\bibitem{Dvo14}
  M.~Dvornikov,
  \textit{Neutrino interaction with matter in a noninertial frame},
  \textit{JHEP}
  \textbf{10} (2014) 053
  [arXiv:1408.2735].

\bibitem{LanLif94}
  L.D.~Landau and E.M.~Lifshitz,
  \textit{The Classical Theory of Fields},
  Butterworth Heinemann, Amsterdam (1994),
  4th edn., pgs.~329--330.

\bibitem{PirRoyWud96}
  D.~P\'{\i}riz, M.~Roy and J.~Wudka,
  \textit{Neutrino oscillations in strong gravitational fields},
  \textit{Phys. Rev.}
  \textbf{D 54} (1996) 1587
  [hep-ph/9604403].

\bibitem{GriMamMos80}
  A.A.~Grib, S.G.~Mamaev and V.M.~Mostepanenko,
  \textit{Quantum Effects in Intense External Fields: Methods and Results not Related to the Perturbation Theory},
  Atomizdat, Moscow U.S.S.R. (1980),
  pp.~13--15.

\bibitem{Bak12}
  K.~Bakke,
  \textit{Noninertial effects on the Dirac oscillator in a topological defect spacetime}, \textit{Eur. Phys. J. Plus}
  \textbf{127} (2012) 82
  [arXiv:1209.0369].

\bibitem{SchWieGre83}
  P.~Schluter, K.-H.~Wietschorke and W.~Greiner,
  \textit{The Dirac equation on orthogonal coordinate systems: I. The local representation}, \textit{J. Phys. A: Math. Gen.}
  \textbf{16} (1983) 1999.

\bibitem{ItzZub80}
  C.~Itzykson and J.-B.~Zuber,
  \textit{Quantum Field Theory},
  McGraw-Hill, New York U.S.A. (1980),
  pp.~691--696.

\bibitem{LanJon09}
  S.K.~Lander and D.I.~Jones,
  \textit{Magnetic fields in axisymmetric neutron stars},
  \textit{Mon. Not. R. Astron. Soc.}
  \textbf{395} (2009) 2162
  [arXiv:0903.0827].

\bibitem{ChaZhi10}
  J.~Charbonneau and A.~Zhitnitsky,
  \textit{Topological currents in neutron stars: kicks, precession, toroidal fields, and magnetic helicity},
  \textit{JCAP}
  \textbf{08} (2010) 010
  [arXiv:0903.4450].

\bibitem{PhiKul94}
  E.S.~Phinney and S.R.~Kulkarni,
  \textit{Binary and millisecond pulsars},
  \textit{Annu. Rev. Astron. Astrophys.}
  \textbf{32} (1994) 591.

\bibitem{Web05}
  F.~Weber,
  \textit{Strange quark matter and compact stars},
  \textit{Prog. Part. Nucl. Phys.}
  \textbf{54} (2005) 193
  [astro-ph/0407155].

\bibitem{MenProMen06}
  D.P.~Menezes, C.~Provid\^{e}ncia and D.B.~Melrose,
  \textit{Quark stars within relativistic models},
  \textit{ J. Phys. G: Nucl. Part. Phys.}
  \textbf{32} (2006) 1081
  [astro-ph/0507529].

\bibitem{NgCheChu03}
  C.Y.~Ng, K.S.~Cheng and M.C.~Chu,
  \textit{Cooling properties of Cloudy Bag strange stars},
  \textit{Astropart. Phys.}
  \textbf{19} (2003) 171
  [astro-ph/0209016].

\bibitem{GriSavStu07}
  A.V.~Grigoriev, A.M.~Savochkin and A.I.~Studenikin,
  \textit{Quantum states of the neutrino in a nonuniformly moving medium},
  \textit{Russ. Phys. J.}
  \textbf{50} (2007) 845.

\bibitem{Dvo14b}
  M.~Dvornikov,
  \textit{Impossibility of the strong magnetic fields generation in an electron-positron plasma},
  \textit{Phys. Rev.}
  \textbf{D 90} (2014) 041702
  [arXiv:1405.3059].

\bibitem{Ber99}
  M.A.~Berger,
  \textit{Introduction to magnetic helicity},
  \textit{Plasma Phys. Control. Fusion}
  \textbf{41} (1999) B167.

\bibitem{DvoSem15JCAP}
  M.~Dvornikov and V.B.~Semikoz,
  \textit{Generation of the magnetic helicity in a neutron star driven by the electroweak electron-nucleon interaction},
  to be published in \textit{JCAP} (2015)
  [arXiv:1503.04162].

\bibitem{HarTad75}
  E.R.~Harrison and E.~Tademaru,
  \textit{Acceleration of pulsars by asymmetric radiation},
  \textit{Astrophys. J.}
  \textbf{201} (1975) 447.

\bibitem{Bur07}
  A.~Burrows, E.~Livne, L.~Dessart, C.D.~Ott and J.~Murphy,
  \textit{Features of the acoustic mechanism of core-collapse supernova explosions},
  \textit{Astrophys. J.}
  \textbf{655} (2007) 416
  [astro-ph/0610175].

\bibitem{KusSeg96}
  A.~Kusenko and G.~Segr\`{e},
  \textit{Velocities of pulsars and neutrino oscillations},
  \textit{Phys. Rev. Lett.}
  \textbf{77} (1996) 4872
  [hep-ph/9606428].

\bibitem{Tot98}
  T.~Totani, K.~Sato, H.E.~Dalhed and J.R.~Wilson,
  \textit{Future detection of supernova neutrino burst and explosion mechanism},
  \textit{Astrophys. J.}
  \textbf{496} (1998) 216
  [astro-ph/9710203].

\bibitem{KeiRafJan03}
  M.T.~Keil, G.G.~Raffelt and H.-T.~Janka,
  \textit{Monte Carlo study of supernova neutrino spectra formation},
  \textit{Astrophys. J.}
  \textbf{590} (2003) 971
  [astro-ph/0208035].

\bibitem{Yak11}
  D.G.~Yakovlev, W.C.G.~Ho, P.S.~Sternin, C.O.~Heinke and A.Y.~Potekhin,
  \textit{Cooling rates of neutron stars and the young neutron star
  in the Cassiopeia A supernova remnant},
  \textit{Mon. Not. Roy. Astron. Soc.}
  \textbf{411} (2011) 1977
  [arXive:1010.1154].

\bibitem{HehNi90}
  F.W.~Hehl and W.-T.~Ni,
  \textit{Inertial effects of a Dirac particle},
  \textit{Phys. Rev.}
  \textbf{D 42} (1990) 2045.

\bibitem{BakFur10}
  K.~Bakke and C.~Furtado,
  \textit{Bound states for neutral particles in a rotating frame in the cosmic string spacetime},
  \textit{Phys. Rev.}
  \textbf{D 82} (2010) 2045.

\bibitem{AhmErm02}
  B.J.~Ahmedov,
  \textit{General relativistic galvano-gravitomagnetic effect in current carrying conductors},
  \textit{Phys. Lett.}
  \textbf{A 256} (1999) 9
  [gr-qc/0701045].

\bibitem{StuTok14}
  A.I.~Studenikin and I.V.~Tokarev,
  \textit{Millicharged neutrino with anomalous magnetic moment in rotating magnetized matter},
  \textit{Nucl. Phys.}
  \textbf{B 884} (2014) 396
  [arXiv:1209.3245].

\bibitem{HoyNisBan11}
  C.~Hoyos, T.~Nishioka, and A.~O'Bannon,
  \textit{A chiral magnetic effect from AdS/CFT with flavor},
  \textit{JHEP}
  \textbf{10} (2011) 084
  [arXiv:1106.4030].

\bibitem{GraKapRed14}
  D.~Grabowska, D.~Kaplan and S.~Reddy,
  \textit{The role of the electron mass in damping chiral magnetic instability
  in supernova and neutron stars},
  \textit{Phys. Rev.}
  \textbf{D 91} (2015) 085035
  [arXiv:1409.3602].

\bibitem{SokTer74}
  A.A.~Sokolov and I.M.~Ternov,
  \textit{Relativistic Electron},
  Nauka, Moscow U.S.S.R. (1974),
  p.~354.

\end{thebibliography}
\end{document}